\documentclass[twocolumn, tighten, times]{aastex631}
\usepackage{amsmath}

\shorttitle{FUV-MIR R(V) Relationship}
\shortauthors{Gordon et al.}

\newcommand{\av}{$A(V)$}
\newcommand{\rv}{$R(V)$}
\newcommand{\irv}{$R(V)^{-1} - 3.1^{-1}$}
\newcommand{\alav}{$A(\lambda)/A(V)$}
\newcommand{\elvebv}{$E(\lambda - V)/E(B - V)$}
\newcommand{\fbump}{2175~\AA}

\begin{document}

\title{One Relation for All Wavelengths: \\ The Far-Ultraviolet to Mid-Infrared Milky Way Spectroscopic R(V) Dependent Dust Extinction Relationship}

\author[0000-0001-5340-6774]{Karl D.\ Gordon}
\affiliation{Space Telescope Science Institute, 3700 San Martin
  Drive, Baltimore, MD, 21218, USA}
\affiliation{Sterrenkundig Observatorium, Universiteit Gent,
  Gent, Belgium}

\author[0000-0002-0141-7436]{Geoffrey~C.~Clayton}
\affiliation{Department of Physics \& Astronomy, Louisiana State University,
  Baton Rouge, LA 70803, USA}

\author[0000-0001-9462-5543]{Marjorie Decleir}
\affiliation{Space Telescope Science Institute, 3700 San Martin
  Drive, Baltimore, MD, 21218, USA}

\author[0000-0002-2371-5477]{E.\ L.\ Fitzpatrick}
\affiliation{Department of Astronomy \& Astrophysics, Villanova University,
800 Lancaster Avenue, Villanova, PA 19085, USA}

\author[0000-0002-9139-2964]{Derck Massa}
\affil{Space Science Institute, 4750 Walnut Street, Suite 205, Boulder, CO
80301, USA}

\author{Karl A.\ Misselt}
\affiliation{Steward Observatory, University of Arizona, Tucson,
  AZ 85721, USA}

\author[0000-0002-9599-310X]{Erik J.\ Tollerud}
\affiliation{Space Telescope Science Institute, 3700 San Martin
  Drive, Baltimore, MD, 21218, USA}

\begin{abstract}
Dust extinction is one of the fundamental measurements of dust grain sizes, compositions, and shapes.
Most of the wavelength dependent variations seen in Milky Way extinction are strongly correlated with the single parameter $R(V) = A(V)/E(B-V)$.
Existing \rv\ dependent extinction relationships use a mixture of spectroscopic and photometry observations, hence do not fully capture all the important dust features nor continuum variations.
Using four existing samples of spectroscopically measured dust extinction curves, we consistently measure the \rv\ dependent extinction relationship spectroscopically from the far-ultraviolet to mid-infrared for the first time.
Linear fits of $A(\lambda)/A(V)$ dependent on \rv\ are done using a method that fully accounts for their significant and correlated uncertainties.
These linear parameters are fit with analytic wavelength dependent functions to determine the smooth \rv\ (2.3--5.6) and wavelength (912~\AA--32~\micron) dependent extinction relationship.
This relationship shows that the far-UV rise, \fbump\ bump, and the three broad optical features are dependent on \rv, but the 10 and 20~\micron\ features are not.
Existing literature relationships show significant deviations compared to this relationship especially in the far-ultraviolet and infrared.
Extinction curves that clearly deviate from this relationship illustrate that this relationship only describes the average behavior versus \rv.
We find tentative evidence that the relationship may not be linear with $R(V)^{-1}$ especially in the ultraviolet.
For the first time, this relationship provides measurements of dust extinction that spectroscopically resolve the continuum and features in the ultraviolet, optical, and infrared as a function of \rv\ enabling detailed studies of dust grains properties and full spectroscopic accounting for the effects of dust extinction on astrophysical objects.
\end{abstract}

\section{Introduction} \label{sec:intro}

Dust grains are important in a number of astrophysical processes from radiative transfer throughout a galaxy \citep{Steinacker13} to providing surfaces for efficient $H_2$ formation \citep{Hollenbach71, Wakelam17}.
One of the prime observables for dust grain studies is the extinction of light (absorption and scattering out of the line-of-sight) from single stars due to dust between the observer and the star.
Dust extinction measurements in the ultraviolet (UV, 0.0912--0.3~\micron), optical (0.3--1~\micron), near-infrared (NIR, 1--5~\micron) and mid-infrared (MIR, 5-30~\micron) have shown that dust grains extinguish over all these wavelengths with varying efficiency.
This requires dust grains with sizes from roughly \AA\ to \micron\ \citep{Mathis77, Desert90}.
In addition, such extinction measurements have revealed features that are attributed to carbonaceous and silicate grains.
The carbonaceous features are the \fbump\ bump \citep{Stecher65, Fitzpatrick86}, the 3.4~\micron\ aliphatic feature \citep{Soifer76, Pendleton02}, and the weak 3.3, 6.2, and 7.7~\micron\ aromatic features \citep{Hensley20, Potapov20}.
The silicate features are the strong 10 and 20~\micron\ features \citep{Roche84, Rieke85, Schutte98, Kemper04, Gordon21}.
Thus, the overall wavelength dependent extinction behavior and the details of these spectroscopic features provide important quantitative constraints on the size, composition, and shape of dust grains.
Other dust observables that are also crucial to constraining dust grain properties include optical and infrared emissions, extinction and emission polarizations, and dust atomic abundances from depletion measurements \citep {Li97, Weingartner01, Zubko04, Jones13, Siebenmorgen14}.

Dust extinction curves in the Milky Way show a range of wavelength-dependent shapes with the largest variations seen in the UV \citep{Massa83, Witt84, Valencic04, Fitzpatrick07}.
The majority of these shape variations are correlated with the ratio of total-to-selective extinction $R(V) = A(V)/E(B-V)$ and this motivated \citet{Cardelli89} to present the first \rv\ dependent extinction relationship to describe Milky Way extinction curves.
There are real deviations from the \rv\ correlations \citep{Mathis92, Clayton03UVDIBs, Valencic04} illustrating that the \rv\ dependent extinction relationship gives the {\em average} variation with \rv.
Even larger deviations from the Milky Way \rv\ dependent extinction relationship can be seen in measurements of dust extinction curves in the Large and Small Magellanic Clouds (LMC and SMC) \citep{Gordon03}\footnote{See \citet{Zagury07} and \citet{Gordon16} for two parameter relationships that can describe MW, LMC, and SMC extinction curves.}.
Nevertheless, finding that the majority of the sightline-to-sightline variation with wavelength in Milky Way extinction curves can be described by a single parameter is an important result.
This means that variations in the dust grains responsible for UV extinction are correlated with the different dust grains that are responsible for the optical and NIR extinction.
Dust grain modelling has shown that this is likely due to correlated changes in the size distribution of dust grains with increasing \rv\ corresponding to an increase in the average grain size \citep{Weingartner01}.

Since \citet{Cardelli89}, a number of studies have derived similar \rv\ dependent extinction relationships for the Milky Way from larger samples and based similarly on UV spectroscopic and optical/NIR photometric extinction measurements \citep[e.g.,][]{Valencic04, Fitzpatrick07}.
Until recently, our understanding of the \rv\ dependence of dust extinction was at spectroscopic resolution in the UV from 1150 to 3100~\AA\ and only photometric in the optical/NIR from 3100~\AA\ to 2.2~\micron.
Fortunately, more recent studies have expanded our knowledge of spectroscopic extinction curves to include the far-UV from 912 to 1190~\AA\ \citep{Gordon09FUSE}, the optical from 0.3 to 1~\micron\ \citep{Fitzpatrick19}, the MIR from 5 to 32~\micron\ \citep{Gordon21}, and the NIR from 0.8 to 5.5~\micron\ \citep{Decleir22}.
All of these studies probe extinction along sightlines towards OB type stars as they are bright from the FUV to the MIR allowing this full wavelength range to be efficiently probed.
Each of these studies has investigated the \rv\ dependent extinction relationship, but based on different samples, different fitting techniques, and over different wavelength ranges.
Simply combining the different \rv\ relationship results in non-physical jumps in the combined relationship in the transition wavelength regions between the studies.
The motivation for this paper is to investigate and derive a single \rv\ dependent extinction relationship from 912~\AA\ to 32~\micron.
This will provide dust extinction curves for a range on environments (i.e., \rv\ values) that can be used to account for the effects of dust extinction on multi-wavelength observations from the far-UV to the mid-infrared without having to use heterogeneous and disjoint relationships for different wavelengths.
For dust studies, a fully spectroscopic \rv\ dependent extinction relationship will enhance studies of the specific grains responsible in a range of environments by providing a comprehensive description of dust extinction from the FUV to the MIR.

In section~\ref{sec:data} we give the details of the four extinction curve samples used.
The derivation of the \rv\ dependent extinction relationship is outlined in section~\ref{sec:results}.
Comparison to previous \rv\ relationships, examples of extinction curves that deviate from the relationship, and a discussion of possible non-linearities with $R(V)^{-1}$ are given in section~\ref{sec:discussion}.
Finally, we give a summary in section~\ref{sec:summary}.

\section{Data \label{sec:data}}

\begin{deluxetable*}{cccccc}[tbp]
\tablewidth{0pt}
\tablecaption{Sample Details\label{tab:samples}}
\tablehead{
  \colhead{Sample} & \colhead{Spectroscopic $\lambda$ [\micron]} & \colhead{\# Sightlines} & \colhead{R(V)} & \colhead{A(V)}
  & \colhead{Ref}}
\startdata
GCC09 & 0.912--0.1190, 0.1150--0.31 & 75 & 2.6--5.6 & 0.7--3.2 & \citet{Gordon09FUSE} \\
F19 & 0.1150--0.31, 0.3--0.95 & 69 & 2.5--5.7 & 0.4--4.5 & \citet{Fitzpatrick19} \\
G21 & 0.1150--0.31, 5--32 & 14 & 2.3--4.3 & 1.8--4.6 & \citet{Gordon21} \\
D22 & 0.1150--0.31, 0.8--4.0 & 13 & 2.4--5.3 & 0.8--3.9 & \citet{Decleir22} \\
\enddata
\end{deluxetable*}

The four samples we use are listed in Table~\ref{tab:samples} along with some of their properties.
The combination of these four samples provides full spectroscopic coverage of dust extinction from 912~\AA\ to 32~\micron\ for a broad range of \rv\ values.
All the samples are based on the parent sample of existing International Ultraviolet
Explorer (IUE) spectroscopic extinction curves from 1150--3100~\AA\ \citep{Valencic04, Fitzpatrick07} with the addition of two sightlines observed with the Hubble Space Telescope Imaging Spectrograph (STIS) in the same wavelength range \citep{Clayton03UVDIBs}.
This work focuses on the dust in the diffuse interstellar medium (ISM), hence we have not used sightlines that are known to probe the dense ISM as defined by the presence of the 3~\micron\ ice absorptio feature.
The \citet{Gordon09FUSE} sample consists of 75 extinction curves using Far Ultraviolet Spectroscopic Explorer \citep[FUSE,][]{Sahnow00} spectra from 912--1190~\AA, IUE spectra, and optical/NIR photometry.
The optical region is provided by the \citet{Fitzpatrick19} sample that consists of 69 extinction curves with IUE spectra, STIS spectra from 0.3--0.95~\micron, and NIR photometry.
Three curves were removed from the original sample of 72 due to two of them having very large uncertainties on \rv\ or no JHK measurements and one being of a known dense sightline (HD~29647).
The \citet{Gordon21} sample of 14 sightlines is based on IUE and STIS spectra, optical/NIR/MIR photometry, and Spitzer Space Telescope Infrared Spectrograph (IRS) spectra from 5--32~\micron.
\citet{Decleir22} gives a sample of 13 sightlines based on IUE and STIS spectra, optical photometry, and NASA Infrared Telescope Facility (IRTF) SpeX spectra from 0.8--5.5~\micron.
For both the G21 and D22 samples, the two known dense sightlines (HD~29647 and HD~283809) were removed.

\begin{figure}[tbp]
\epsscale{1.15}
\plotone{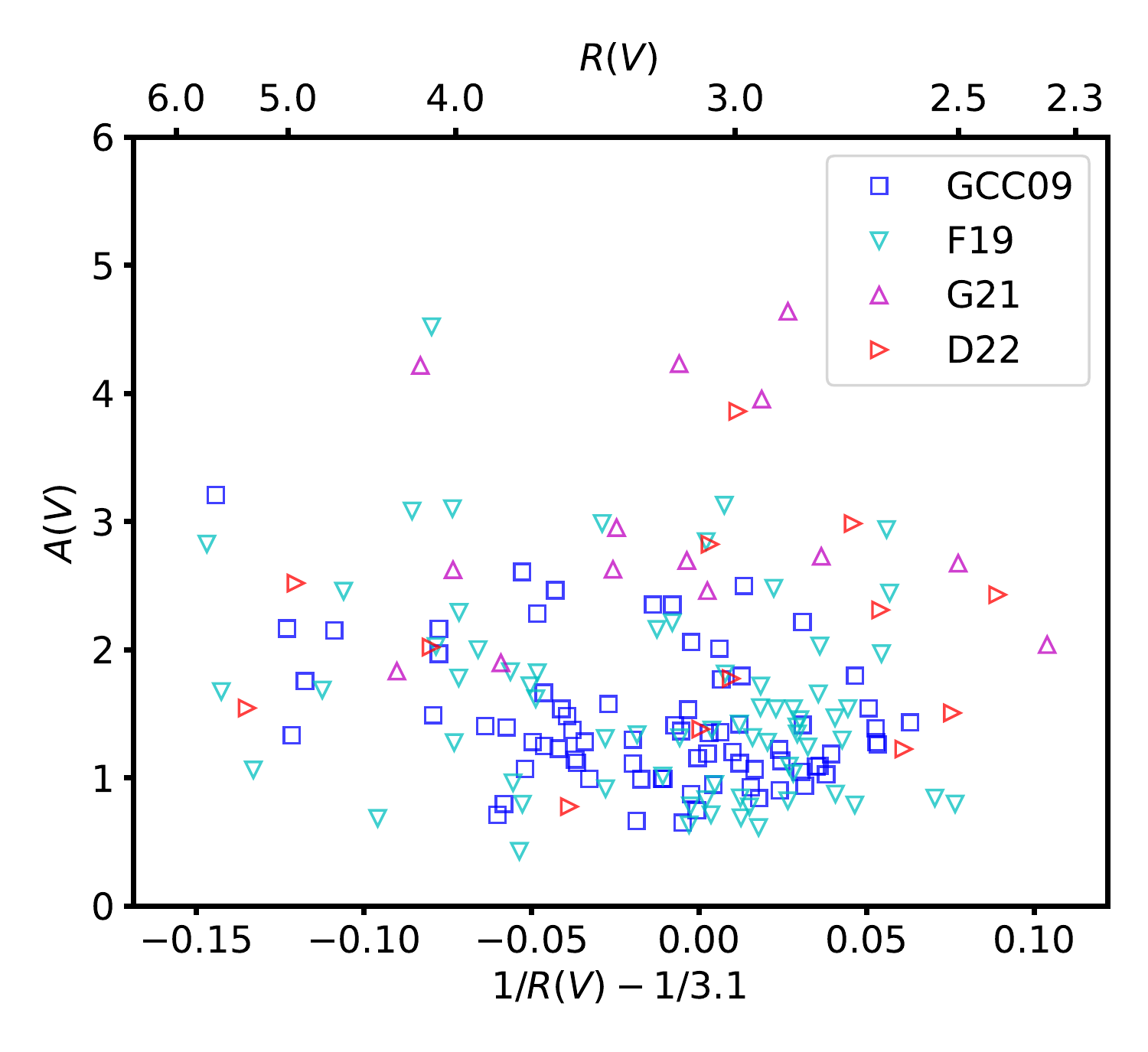}
\caption{The \rv\ and \av\ properties of the different samples are shown.
The samples are GCC09 \citep{Gordon09FUSE}, F19 \citep{Fitzpatrick19}, G21 \citep{Gordon21}, and D22 \citep{Decleir22}.
We plot $1/R(V) - 1/3.1$ instead of \rv\ as this is the quantity used to derive the \rv\ dependent extinction relationship (see sec~\ref{sec:results} for details).
The top x-axis provides the corresponding \rv\ values.
\label{fig:sampprop}}
\end{figure}

The samples from the different studies used here cover different ranges in dust column \av, average grain size \rv, and number of sightlines.
This is illustrated in Fig.~\ref{fig:sampprop} where \av\ is plotted versus \rv.
The \av\ and \rv\ values were determined by extrapolating the photometric extinction curves beyond 1~\micron\ to infinite wavelength assuming a functional form of the extinction curve as is commonly done \citep{Cardelli89, Martin90}.
For the samples with NIR or MIR spectroscopic data, the $>1$~\micron\ spectroscopy and photometry was extrapolated using a power-law plus modified Drude functions to account for the known spectral features \citep{Gordon21, Decleir22}.
For the samples with only JHK photometry, this photometry was extrapolated assuming the \citet{Rieke89} extinction curve \citep{Gordon09FUSE, Fitzpatrick19}.
For the \citet{Fitzpatrick19} sample, we assume $V$ is the same as the measurement at 5500~\AA\ and $B$ at 4400~\AA\ as they have shown this to be a good approximation.

All four samples can be thought of as representative rather than statistical samples.
Representative samples are useful when the goal is to study the dust extinction behavior versus an environmental parameter like \rv.
Such representative studies are complementary to statistical sample based studies where the usual goal is to determine the average behavior.
Two of the samples were the result of targeted observations that were picked to be representative with sightlines distributed as uniformly as possible in \rv\ \citep{Gordon21, Decleir22}.
The other two samples included all possible measurements available in archives \citep{Gordon09FUSE} or using a subsample of the existing IUE extinction curve sample \citep{Fitzpatrick19}.
Fortunately all have good sampling over the observed range in \rv.
The sightlines in the different samples have little overlap except for \citet{Gordon21} and \citet{Decleir22} that have 5 sightlines in common.
The limited overlap between samples is partially due to the different sensitivities of the instruments used and the wide range in wavelength.

\section{Results \label{sec:results}}

\subsection{Linear fits at each wavelength}
\label{ssec:linfits}

Before performing any fits, the spectroscopic data from each sample were corrected for hydrogen atomic (e.g., Ly$\alpha$ at 912~\AA) and molecular (912--1100~\AA) absorption lines including masking wavelength regions where such corrections were not sufficiently accurate \citep[e.g.,][]{Gordon09FUSE}.
The resulting spectra were re-binned to a resolving power of 500 if that was lower than the native resolution.
We found it better to rebin the extinction curves to a reasonable resolution as this gives higher signal-to-noise measurements where the known broad dust features are well resolved.
Narrow extinction features like diffuse interstellar bands (DIBs) are not resolved at the spectral resolutions of these datasets.
The four individual study papers provide details of the analysis at each dataset's full spectral resolution.

The basic fit needed to define the \rv\ relationship is to fit a linear function to extinction versus \rv\ at all measured wavelengths.
For more on why a linear function is assumed, see Section \ref{sec:whylin}.
Extinction can be parameterized as \alav\ or \elvebv, and both have been used to characterize the \rv\ relationship \citep{Cardelli89, Fitzpatrick99review}.
Here, we choose to use \alav\ as it is a measure of the absolute extinction rather than relative as well as being  more directly comparable to dust grain models.
In addition, it is more generally useful for characterizing the foreground extinction towards astrophysical objects.
We note that the \rv\ dependent extinction relationships derived with either formulation are mathematically equivalent \citep[see Appendix B,][]{Fitzpatrick19}.

As we are using \alav, we fit versus \irv.
Specifically
\begin{equation}
\frac{A(\lambda)}{A(V)} = a(\lambda) + b(\lambda) \left[ \frac{1}{R(V)} - \frac{1}{3.1} \right]
\label{eq_fitmod}
\end{equation}
where $a(\lambda)$ is the y-intercept and $b(\lambda)$ is the slope for a given wavelength.
The subtraction of $3.1^{-1}$ from $R(V)^{-1}$ shifts the y-intercept coefficient, $a(\lambda)$, to be the extinction for
$R(V) = 3.1$, the often used Milky Way average value \citep{Johnson65, Schultz75, Whittet80, Fitzpatrick99}.

Note that if we had fit \alav\ versus \rv\ rather than versus a function of $R(V)^{-1}$, we would need to fit a non-linear function as the relationship would be necessarily non-linear.
This can be illustrated by considering the B band relationship where such a fit would be between $A(B)/A(V)$ versus $R(V) = A(V)/E(B-V) = 1 / [A(B)/A(V) - 1]$ and this is clearly not linear.

Performing the linear fit is challenging as both \alav\ and $R(V)^{-1} = E(B-V)/A(V)$ have significant uncertainties and their uncertainties are correlated due to both being normalized by \av.
The covariance for the $i$th point is calculated using
\begin{equation}
\mathrm{cov_i(\lambda)}^2 = \left( \frac{1}{R(V)_i} - \frac{1}{3.1} \right) \left( \frac{A(\lambda)}{A(V)} \right)_i \left( \frac{\sigma[A(V)_i]}{A(V)_i} \right)^2 .
\end{equation}
While this is an approximation to the full covariance calculation, we have found it captures the vast majority of the covariance.
We investigated existing techniques for performing the linear fits including unweighted fits, y uncertainty weighted fits, orthogonal distance regression \citep{Robotham15}, and Monte Carlo sampling.
All of these techniques rely on evaluating the likelihood (e.g., $\chi^2$) at one point on the proposed model.
After applying the existing techniques at all wavelengths, we found that none of them did well at reproducing the average behavior of the measured extinction curves in the far-UV at either high \rv, low \rv\, or both.
As a result, our fitting uses a custom likelihood based on line integrals to account for the full impact of these correlated uncertainties.

\begin{figure}[tbp]
\epsscale{1.1}
\plotone{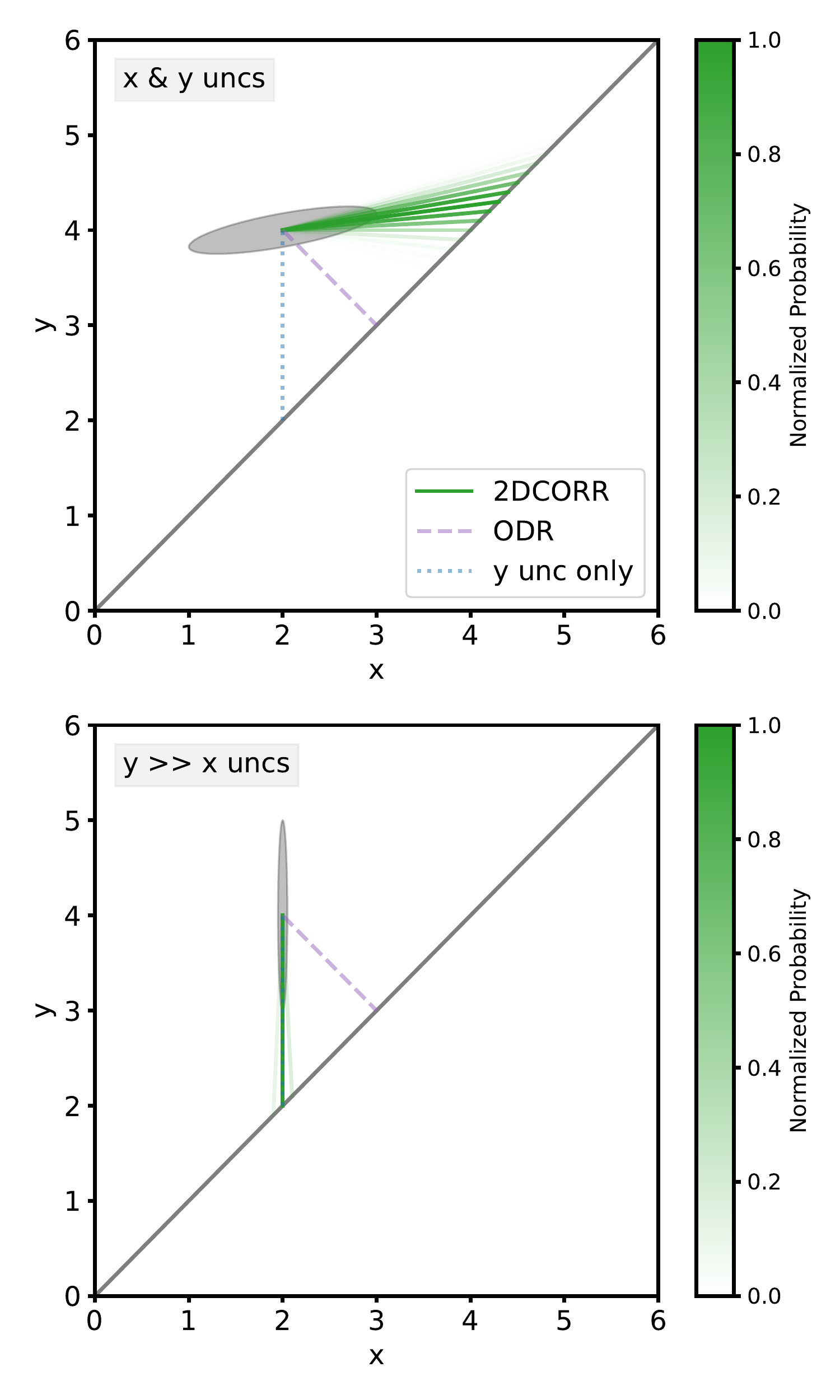}
\caption{The details of the 2DCORR likilihood calculation is illustrated for a single data point with correlated x and y uncertainties (top) and with small x uncertainties (bottom).
The green 2DCORR lines gives the normalized probability that the data point is consistent with the model line at that point on the model line.
While this probability was calculated at many points along the line, many are vanishingly small compared to the maximum probability and hence to not show up in this figure.
For comparison, the point on the model line for standard y uncertainty only and orthogonal distance regression (ODR) methods are illustrated.
Note that the 2DCORR converges to the standard "y unc only" case when the x uncertainties are small.
\label{fig:fit_example}}
\end{figure}

\begin{figure*}[tbp]
\epsscale{1.2}
\plotone{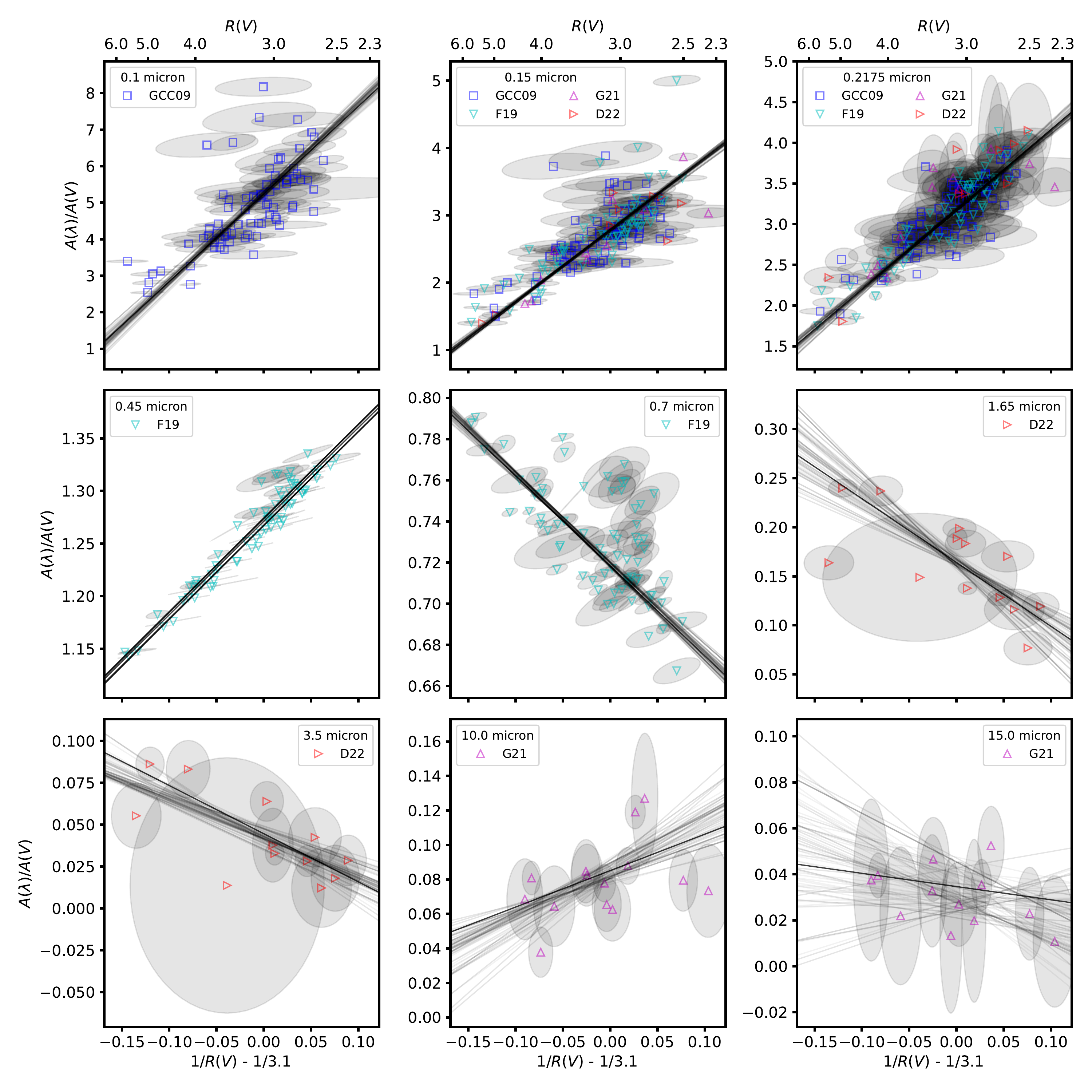}
\caption{The normalized extinction \alav\ is plotted versus \irv\ for a set of representative wavelengths.
The samples are indicated in the legends with the correlated uncertainties shown as gray ellipses.
The 2DCORR linear fits accounting for the correlated uncertainties are shown as solid lines with the fainter lines illustrating the uncertainty in the fits based on random sampling of the MCMC chains.
\label{fig:repwaves}}
\end{figure*}

Our fitting technique, hereby named 2DCORR\footnote{We are certain this technique has been published before, but cannot find a reference.  We await notification of such a reference, likely written many decades ago.}, relies on computing a likelihood function that allows for non-symmetric and different-for-all-points (i.e., heteroskedastic) uncertainties for all extinction measurements at a wavelength $\lambda$.
This likelihood is based on the assumption that the data are drawn from a distribution composed of the model line (Equation~\ref{eq_fitmod}) plus a 2D multivariate Gaussian measurement uncertainty (the two dimensions are \alav\ and \irv).
Convolving these distributions and recasting the problem as a line integral along the model line yields the likelihood inside the product in Eq.~\ref{eq_like}.
The 2DCORR line integral calculation is illustrated in Fig.~\ref{fig:fit_example}.
This accounts for the probability that the observed data point at that wavelength could have originated from any point along the model line and not just from a single point as is assumed in the techniques listed in the previous paragraph.
The full likelihood for each wavelength bin is then the product of these individual likelihoods for all the datasets with measurements in that wavelength bin.
Thus
\begin{equation}
 \mathcal{L} (a, b) = \prod_{i=1}^n \frac{\int_{l_0}^{l_1} \mathcal{N}(\mu_i, \Sigma_i) dl}
                                  {\int_{l_0}^{l_1} dl}
\label{eq_like}
\end{equation}
where the product is over the $n$ data points at that wavelength, $\mathcal{N}$ is the 2D Gaussian evaluated at that point on the line for the $i$th data point, $\mu_i = (x_i, y_i)$, $x_i = R(V)_i^{-1} - 3.1^{-1}$, $y_i = A(\lambda)/A(V)_i$, $\Sigma_i$ is the corresponding $i$th covariance matrix, and the integral is done along the the model line $l$ defined as Equation~\ref{eq_fitmod}.
The integral is calculated along the model line between min/max values picked to encompass a range along the line that includes all the data points and sufficient multiples of their uncertainties.
In practice, the line integral is done numerically by dividing the specified $x$ range into many points, evaluating the model line at all these points and then using the resulting $(x, y)$ pairs to numerically integrate $\mathcal(N)$ along the model line.
The denominator provides the normalization by the line length as the line integral is done over min/max $x$ values and, thus, over different line lengths for different values of $a$ and $b$.

To infer the posterior probability of the $a$ and $b$ parameters for each wavelength bin, we assume uniform priors, and follow a two step process to sample the posterior.
First, we find a maximum a posteriori estimate of the solution (in practice due to our uniform priors, a maximum-likelihood estimate) using iterative sigma rejection (3 times with a $5\sigma$ threshold) using the `LevMarLSQFitter' from the astropy modeling package \citep{astropy:2022} maximizing the likelihood given by Eq.~\ref{eq_like}.
We then used the maximum likelihoods to provide initial values for sampling the posterior probability for the intercept ($a$) and slope ($b$) at each wavelength.
We sample the posterior  using the `emcee' Monte Carlo Markov Chain package \citep{emcee}, and determine stopping points for the sampling by inspection of the chains.
We then summarize the posterior distributions from the MCMC chains as the mean and standard deviation of the samples for $a$ and $b$ independently.

Example fits are shown in Fig.~\ref{fig:repwaves} sampling a wide range of wavelengths.
Only spectroscopic extinction curve data were used for the fits.
The data points have ellipses showing the correlated x and y 1-$\sigma$ uncertainties.
The importance of accounting for the full 2D uncertainties is illustrated in the far-UV (0.1~$\micron$,upper-left panel of Fig.~\ref{fig:repwaves}) where there are four data points at high \alav\ values that would be significantly deviant from the displayed best fit line if only $y$ uncertainties were used, but are consistent with the fit when including the relatively large $x$ uncertainties.

For the UV wavelengths where multiple samples have spectroscopic data, the fits are shown to the combined sample.
The linear fits using the 2DCORR technique are shown as solid lines.
The fits are a good representation of the data and uncertainties, with only a few points being significantly deviant from the fitted line.

\subsection{R(V) relationship}

The linear fit parameters at each wavelength combine to give the \rv\ relationship as a function of wavelength where the intercept gives the extinction curve at $R(V) = 3.1$ and the slope gives the variations as a function of \rv.
In the following three subsections, we show the detailed behavior and fits to typical analytic extinction curve models in the UV, optical, and NIR/MIR wavelength regions, because they each show qualitatively different behavior.
Finally, the behavior combined over the entire wavelength range is investigated.
As has been done for most previous \rv\ relationships \citep[e.g.,][]{Cardelli89, Fitzpatrick99review}, the analytic fits provide a relationship that is contiguous and smoothly interpolates over any spectral gaps.
All the analytic best-fit models are obtained from maximum-likelihood fitting to the mean of the posterior distributions from Section \ref{ssec:linfits} using the `LevMarLSQFitter' from the astropy modeling package \citep{astropy:2022}, supplemented with iterative $\sigma$ rejection (3 times with a $5\sigma$ threshold).
We did not use the uncertainties on linear fits parameters as relative weights as this caused the fit to visually deviate from some of the data in the overlap regions between datasets (e.g, around 0.9~\micron\ for F19 and D22).  This is likely due to the differences in magnitude of the uncertainties between different observation samples.

\subsubsection{Ultraviolet}
\label{sec:rv_uv}

\begin{figure*}[tbp]
\epsscale{1.1}
\plotone{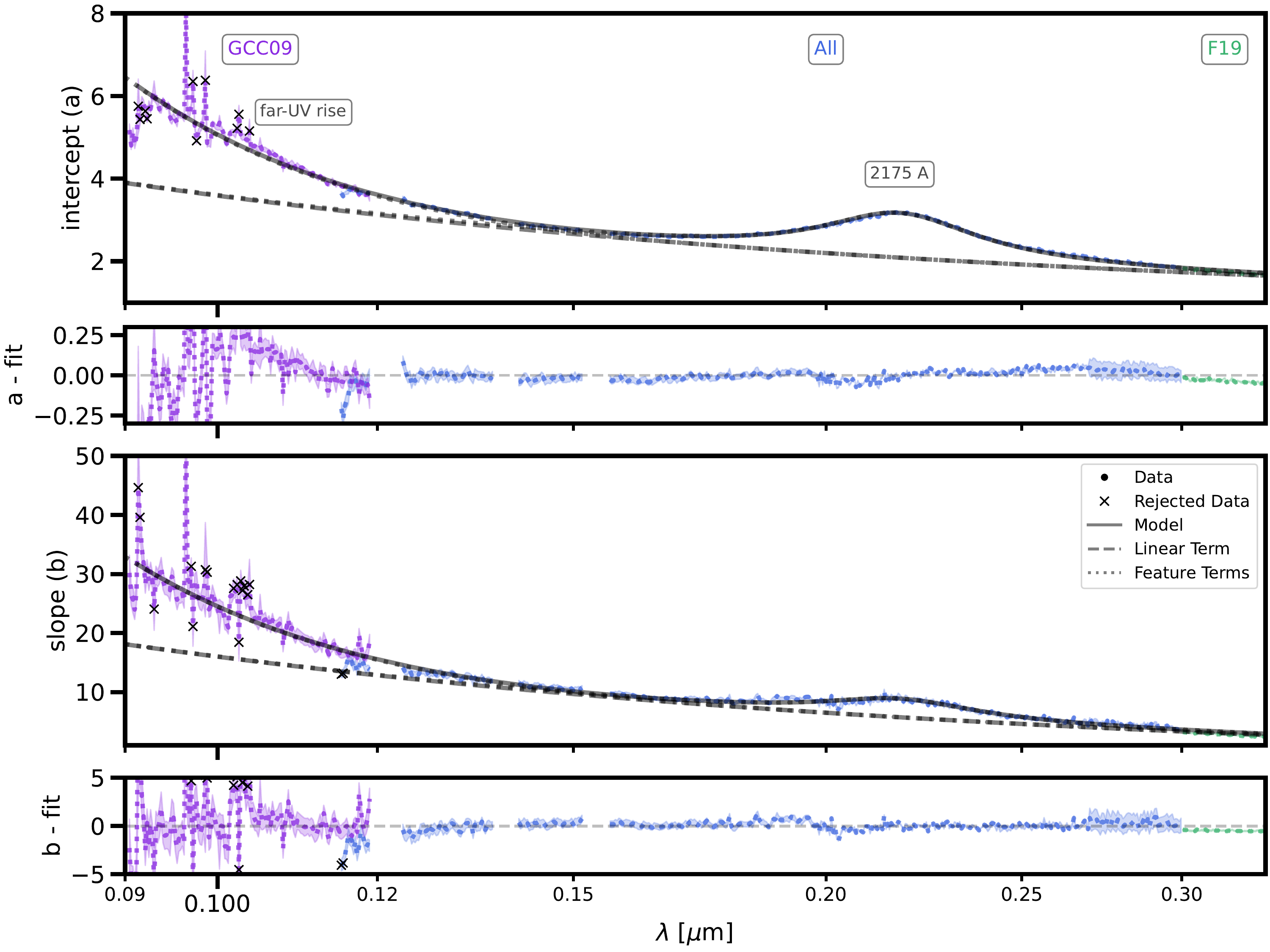}
\caption{Linear fit parameters for UV wavelengths are plotted where the shaded regions give the fit parameter uncertainties.
The results labeled as ``All'' are from fits to the combined GCC09, F19, D22, and G21 extinction curves at UV wavelengths (0.115--0.3~\micron).
Wavelengths with known issues with measuring extinction (e.g., stellar wind lines) have not been fit.
The residuals from FM90 fits to the intercept and slope values are shown in the two narrow plots.
\label{fig:uvwaves}}
\end{figure*}

\begin{deluxetable}{lDD}
\tablewidth{0pt}
\tablecaption{UV Parameters\label{tab:uvparam}}
\tablehead{\colhead{parameter} & \twocolhead{$a_\mathrm{uv}$} & \twocolhead{$b_\mathrm{uv}$} }
\startdata
\decimals
$c^{AV}_1$ & 0.81297 & -2.97868 \\
$c^{AV}_2$ & 0.2775 & 1.89808 \\
$c^{AV}_3$ & 1.06295 & 3.10334 \\
$c^{AV}_4$ & 0.11303 & 0.65484 \\
$x_o$      & \multicolumn{4}{c}{4.60~\micron$^{-1}$ = 2174~\AA} \\
$\gamma$   & \multicolumn{4}{c}{0.99~\micron$^{-1}$ = 49~\AA} \\
\enddata
\end{deluxetable}

The wavelength dependence of the intercepts and slopes in the UV wavelength region are plotted in Fig.~\ref{fig:uvwaves}.
UV extinction curves are often fit using the FM90 parametrization \citep{Fitzpatrick90}.
The FM90 formulation is a combination of a line, a Drude for the \fbump\ bump, and a quadratic for the far-UV rise.
We have found the FM90 parameterization fits the wavelength dependence of both coefficients.
Specifically,
\begin{equation}
\label{eq:uv}
\left.
\begin{matrix}
a_\mathrm{uv}(x) \\
b_\mathrm{uv}(x)
\end{matrix}
\right\} =
   c_1^{AV} + c_2^{AV}x + c_3^{AV}D(x, x_o, \gamma) + c_4^{AV}F(x)
\end{equation}
where $x = 1/\lambda~[\micron^{-1}]$, the \fbump\ bump is
represented by the Drude term
\begin{equation}
\label{eq:drude}
D(x, x_o, \gamma) = \frac{x^2}{(x^2 - x_o^2)^2 + (x\gamma)^2},
\end{equation}
where $x_o$ is the center and $\gamma$ is the width.
The far-UV curvature (for $x \ge 5.9~\micron^{-1}$, $\lambda < 0.1695~\micron$) is given by
\begin{equation}
F(x) = 0.5392(x - 5.9)^2 + 0.05644(x - 5.9)^3.
\end{equation}
Note that the central amplitude of the \fbump\ bump is given as $c^{AV}_3 / \gamma^2$.

While the FM90 parametrization is usually applied to extinction curves, it has been empirically shown that it also fits the intercept and slope of \rv\ dependent extinction relationships in the UV \citep{Cardelli89}.
\citet{Gordon09FUSE} showed that the FM90 formula works over the entire UV wavelength range all the way to 912~\AA.
We used the FM90 model from the `dust\_extinction' python package \citep{dustextinction} to fit to the intercept and slope values from 912--3300~\AA\ for the GCC09, All, and F19 samples.
Following \citet{Fitzpatrick19}, we give five times lower weight to the longest wavelengths (2700--3000~\AA) in the ``All" sample due to issues with the IUE spectra at these wavelengths.
The FM90 fit parameters are given in Table~\ref{tab:uvparam}.
The FM90 fits and residuals are plotted in Fig.~\ref{fig:uvwaves} illustrating that the fits do a good job of describing the wavelength dependence of the intercept and slope parameters.
As expected due to the IUE issue at longer UV wavelengths, the residuals to the fit are elevated between 2700--3300~\AA.
In addition, there is structure in the residuals between 1000--1100~\AA\ for the intercepts, but not the slopes.
This residual structure may be due to small errors in the $H_2$ model used to correct for the $H_2$ absorption prevalent in this wavelength range \citep{Gordon09FUSE}.
The \fbump\ bump is clearly elevated in the slopes when compared to nearby wavelengths and the far-UV rise component is elevated above the linear trend indicating these features are grain size dependent.

\subsubsection{Optical}
\label{sec:rv_opt}

\begin{figure*}[tbp]
\epsscale{1.1}
\plotone{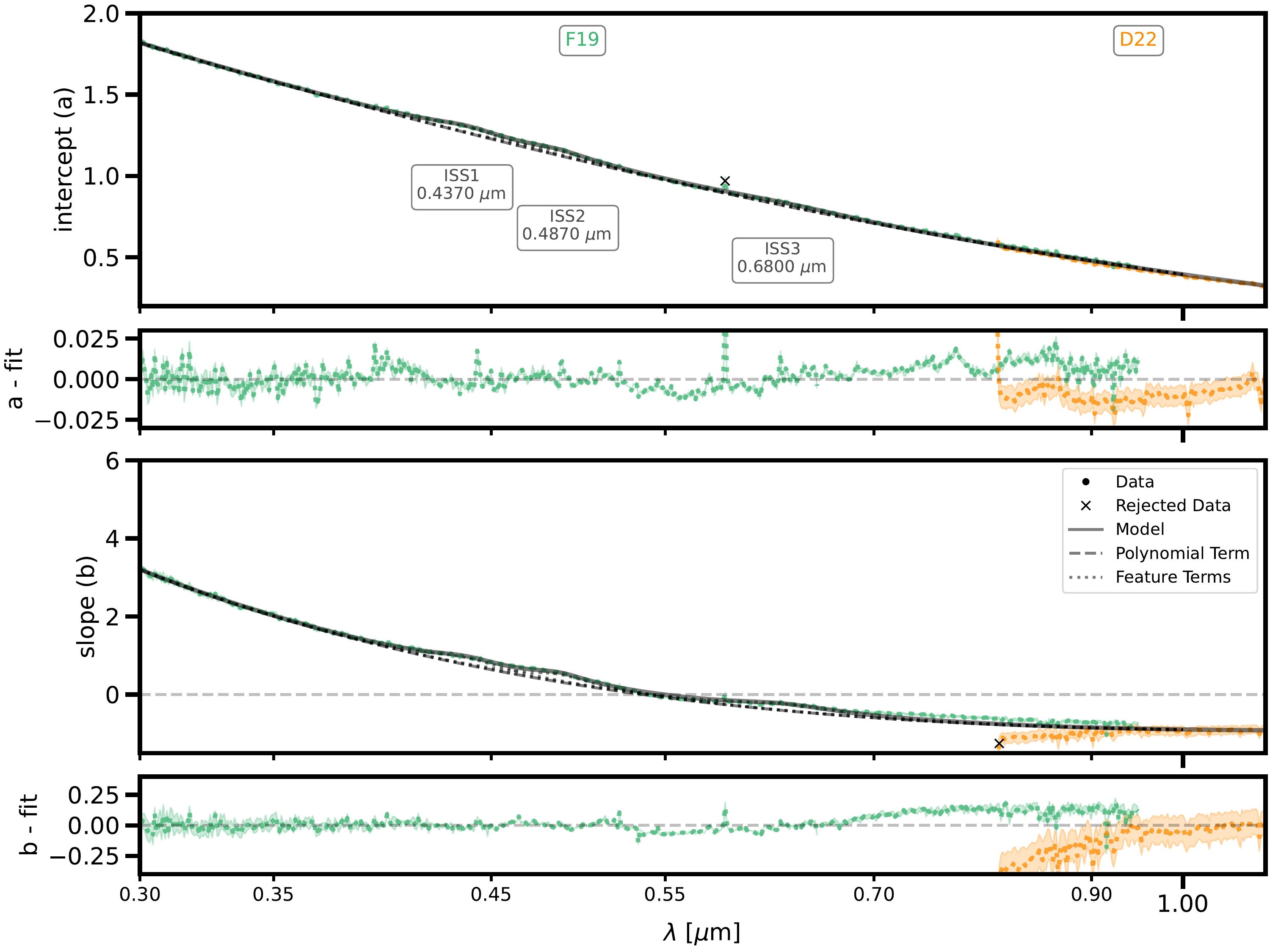}
\caption{Linear fit parameters for optical wavelengths are plotted where the shaded regions give the fit parameter uncertainties.
Wavelengths with known issues with measuring extinction (e.g., H$\alpha$) have not been fit.
The residuals from fits to the intercept and slope values are shown in the two narrow plots.
\label{fig:optwaves}}
\end{figure*}

\begin{deluxetable}{lDD}
\tablewidth{0pt}
\tablecaption{Optical Parameters\label{tab:optparam}}
\tablehead{\colhead{parameter} & \twocolhead{$a_\mathrm{opt}$} & \twocolhead{$b_\mathrm{opt}$}}
\startdata
\decimals
$E_0$ & -0.35848 & 0.12354 \\
$E_1$ & 0.7122 & -2.68335 \\
$E_2$ & 0.08746 & 2.01901 \\
$E_3$ & -0.05403 & -0.39299 \\
$E_4$ & 0.00674 & 0.03355 \\
$F_1$ & 0.03893 & 0.18453 \\
$x_1$ & \multicolumn{4}{c}{2.288~\micron$^{-1}$ = 4371~\AA} \\
$\gamma_1$ & \multicolumn{4}{c}{0.243~\micron$^{-1}$ = 47~\AA} \\
$F_2$ & 0.02965 & 0.19728 \\
$x_2$ & \multicolumn{4}{c}{2.054~\micron$^{-1}$ = 4869~\AA} \\
$\gamma_2$ & \multicolumn{4}{c}{0.179~\micron$^{-1}$ = 43~\AA} \\
$F_3$ & 0.01747 & 0.1713 \\
$x_3$ & \multicolumn{4}{c}{1.587~\micron$^{-1}$ = 6301~\AA} \\
$\gamma_3$ & \multicolumn{4}{c}{0.243~\micron$^{-1}$ = 99~\AA} \\
\enddata
\end{deluxetable}

The wavelength dependence of the intercepts and slopes for the optical wavelength region are plotted in Fig.~\ref{fig:optwaves}.
\citet{Massa20} found that the optical extinction can be well fit with a combination of a low order polynomial for the continuum and three Drude profiles for the Intermediate Scale Structure (ISS) features.
We have found this parameterization fits the wavelength dependence of both coefficients.
Specifically,
\begin{equation}
\label{eq:opt}
\left.
\begin{matrix}
a_\mathrm{opt}(x) \\
b_\mathrm{opt}(x)
\end{matrix}
\right\} =
   \sum_{j=0}^4 E_j x^j + \sum_{i=1}^3 F_i D(x, x_i, \gamma_i) \gamma_i^2
\end{equation}
where $E_j$ give the polynomial coefficients, $F_i$ give the central amplitude of the ISS features, $x = 1/\lambda~[\micron^{-1}]$, $x_i$ give the ISS centers, and $\gamma_i$ give the ISS widths.
The $\gamma_i^2$ in the 2nd sum is needed so the $F_i$ values give the central intensity (see Eq.~\ref{eq:drude}.
Based on the success of the FM90 parameterization fitting the UV slopes and intercepts, we investigated if the \citet{Massa20} parameterization worked for the optical slopes and intercepts.
We fit the F19 and D22 samples to the intercept and slope values from 0.3--1.1~\micron\ using an equal weighting of the two samples.
We found that the fits worked well with fixed values for the ISS centers to 2.288, 2.054, and 1.587 $\micron^{-1}$ and widths to 0.243, 0.179, and 0.243~$\micron^{-1}$ as given by \citet{Massa20}.
The fit parameters are given in Table~\ref{tab:optparam}.
These fits and residuals to them are shown in Fig.~\ref{fig:optwaves}.
The ISS features can be seen to be elevated in the slopes compared to nearby wavelengths.
This is similar to what was seen for the \fbump\ bump indicating that the ISS features are dependent on grain size adding further evidence that the ISS features may be related to the \fbump\ bump \citep{Massa20}.

The results for the intercept and slope for the F19 and D22 samples do not agree well in the overlap region between 0.8 and 0.95~\micron.
For the F19 sample, this wavelength region includes the Paschen jump and is known not to be fully included in the stellar atmosphere models used to measure the F19 extinction curves \citep{Fitzpatrick19, Massa20}.
For the D22 sample, observed standards were used instead of stellar atmosphere models and the clear residuals seen for the Paschen jump and lines indicate some spectral type mismatch between the reddened and standard stars \citep{Decleir22}.
Overall these fits provide a good description of the wavelength dependence in the optical region except in the 0.8--0.95~\micron\ range where the fits are intermediate between the two.

\subsubsection{Near- and Mid-Infrared}
\label{sec:rv_ir}

\begin{figure*}[tbp]
\epsscale{1.1}
\plotone{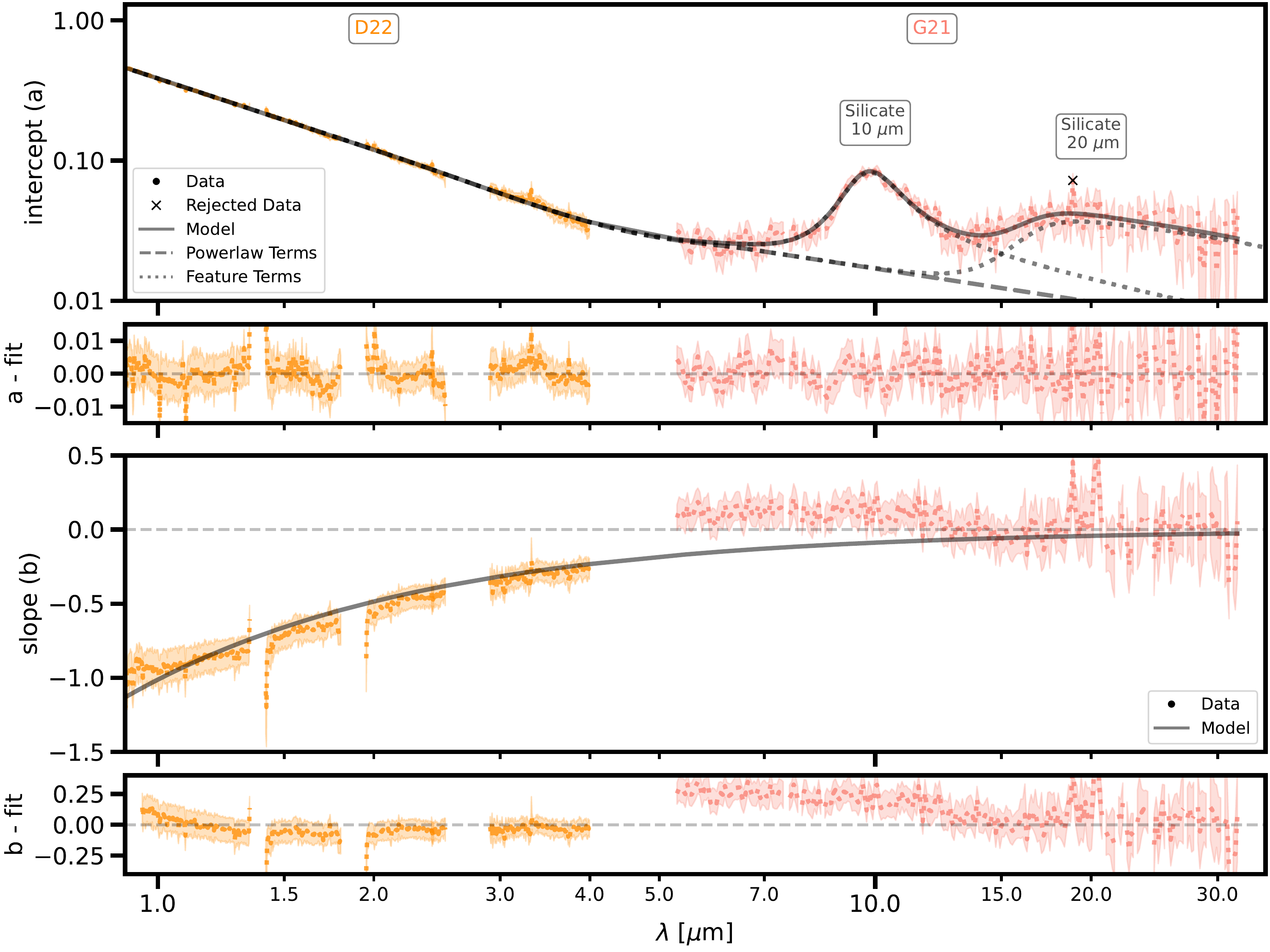}
\caption{Linear fit parameters for NIR and MIR wavelengths are plotted where the shaded regions give the fit parameter uncertainties.
Wavelengths with known issues with measuring extinction (e.g., strong atmospheric absorption) have not been fit.
The residuals from fits to the intercept and slope values are shown in the two narrow plots.
\label{fig:irwaves}}
\end{figure*}

\begin{deluxetable}{lDD}
\tablewidth{0pt}
\tablecaption{NIR/MIR Parameters\label{tab:irparam}}
\tablehead{\colhead{parameter} & \twocolhead{$a_\mathrm{ir}$} & \twocolhead{$b_\mathrm{ir}$} }
\startdata
\decimals
$g_1$           & 0.38526 & -1.01251 \\
$\alpha_1$      & 1.68467 & -1.06099 \\
$\alpha_2$      & 0.78791 \\
$\lambda_b$     & 4.30578 \,\micron \\
$\delta$        & 4.78338 \,\micron \\
$S_1$           & 0.06652 \\
$\lambda_{o,1}$ & 9.8434 \,\micron \\
$\gamma_{o_1}$  & 2.21205 \,\micron \\
$a_1$           & -0.24703 \\
$S_2$           & 0.0267 \\
$\lambda_{o,2}$ & 19.258294 \,\micron \\
$\gamma_{o_2}$  & 17.0 \,\micron \\
$a_2$           & -0.27 \\
\enddata
\end{deluxetable}

The wavelength dependence of the intercepts and slopes for the NIR and MIR wavelength regions are plotted in Fig.~\ref{fig:irwaves}.
In contrast to the \fbump\ and ISS features, the 10 and 20~\micron\ silicate features are clearly seen in the intercepts, but not in the slope values.
Note that there were not enough NIR extinction curve measured between 4 and 5.5~\micron\ to derive the intercept and slope values.
\citet{Gordon21} found that the MIR spectroscopic extinction can be well fit with a power law and two modified Drude profiles for the two silicate features.
\citet{Decleir22} found that the NIR spectroscopic extinction can be well fit with a power law with the addition of a modified Drude profile for the 3~\micron\ ice feature for dense sightlines.
We do not include dense sightlines in this work, so do not include a Drude for the 3~\micron\ ice feature in our fitting.
In fitting the NIR and MIR intercepts, we found that a single power law plus two modified Drude profiles were not sufficient.
Fitting such a function clearly indicated that two power laws were needed with one power law for a portion of the NIR and another for the rest of the MIR.
Such a result is consistent with the \citet{Gordon21} and \citet{Decleir22} work as they found different power law indices.

The functional form used for fitting the intercepts is
\begin{eqnarray}
\label{eq:ir1}
a_\mathrm{ir}(\lambda) & = & g_1 \lambda^{-\alpha_1} \left[1 - W(\lambda, \lambda_b, \delta) \right] \\
   & & + g_1 \frac{\lambda_b^{-\alpha_1}}{\lambda_b^{-\alpha_2}} \lambda^{-\alpha_2} W(\lambda, \lambda_b, \delta) \nonumber \\
   & & + \sum_{i=1}^2 S_i D_m(\lambda, \lambda_{o,i}, \gamma_{o,i}, a_i)
\end{eqnarray}
where $\alpha_i$ are the power law indices, $\lambda_b$ is the break wavelength between the two power laws, and $\delta$ is the width for the transition between the power laws.
The transition between the two power laws is given by the smooth step function
\begin{equation}
\label{eq:smoothstep}
W(\lambda, \lambda_b, \delta) =
    \left\{ \begin{array}{ll}
        0 & \mbox{if $z < 0$}; \\
        3z^2 - 2z^3 & \mbox{if $0 \leq z \leq 1$}; \\
        0 & \mbox{if $z > 1$}.
           \end{array}
    \right.
\end{equation}
where $\lambda_b$ is the break wavelength, $\delta$ is the width, and their relationship with $z$ is
\begin{equation}
z = \frac{\lambda - (\lambda_b - 0.5\delta)}{\delta} .
\end{equation}
The modified Drude profile provides more asymmetric profiles than given by the intrinsic Drude asymmetry and is
\begin{equation}
D_m(\lambda, \lambda_o, \gamma_o, a) = \frac{(\gamma/\lambda_o)^2}{(\lambda / \lambda_o - \lambda_o / \lambda)^2 + (\gamma / \lambda_o)^2} \\
\end{equation}
where
\begin{equation}
\gamma = \frac{2 \gamma_o}{1 + \exp [a (\lambda - \lambda_o)]},
\end{equation}
$\lambda_o$ is the central wavelength, $\gamma_o$ is the unmodified width, and $a$ is the ``extra'' asymmetry parameter.
We fix $\gamma_{o,2} = 17\,\micron$ and $a_2 = -0.27$ following \citet{Gordon21} as these two parameters are not well constrained.
The fit parameters are given in Table~\ref{tab:irparam}.

The functional form needed to fit the slopes is quite different than the intercepts.
Many of the slopes are negative and roughly follow a power law with the exception of the slopes between 5 and 13~\micron\ that are slightly positive with values around 0.1.
There is no indication of the 10 and 20~\micron\ silicate features.
In addition, it is highly likely that the slopes between 1--5~\micron\ are affected by residual atmospheric features \citep{Decleir22} appearing as a majority of the high frequency variations seen.
Given these two points, we use a simple power law for the slopes with
\begin{equation}
\label{eq:ir2}
b_\mathrm{ir}(\lambda) = g_1 \lambda^{-\alpha_1} .
\end{equation}
We tested more complex fitting functions that reproduced the detailed behavior of the slope values, but found them to produce extinction curves that did not reproduce the actual measured curves at high and low \rv\ values.
The fit parameters are given in Table~\ref{tab:irparam}.

\subsubsection{All Wavelengths}
\label{sec:rv_all}

\begin{figure*}[tbp]
\epsscale{1.1}
\plotone{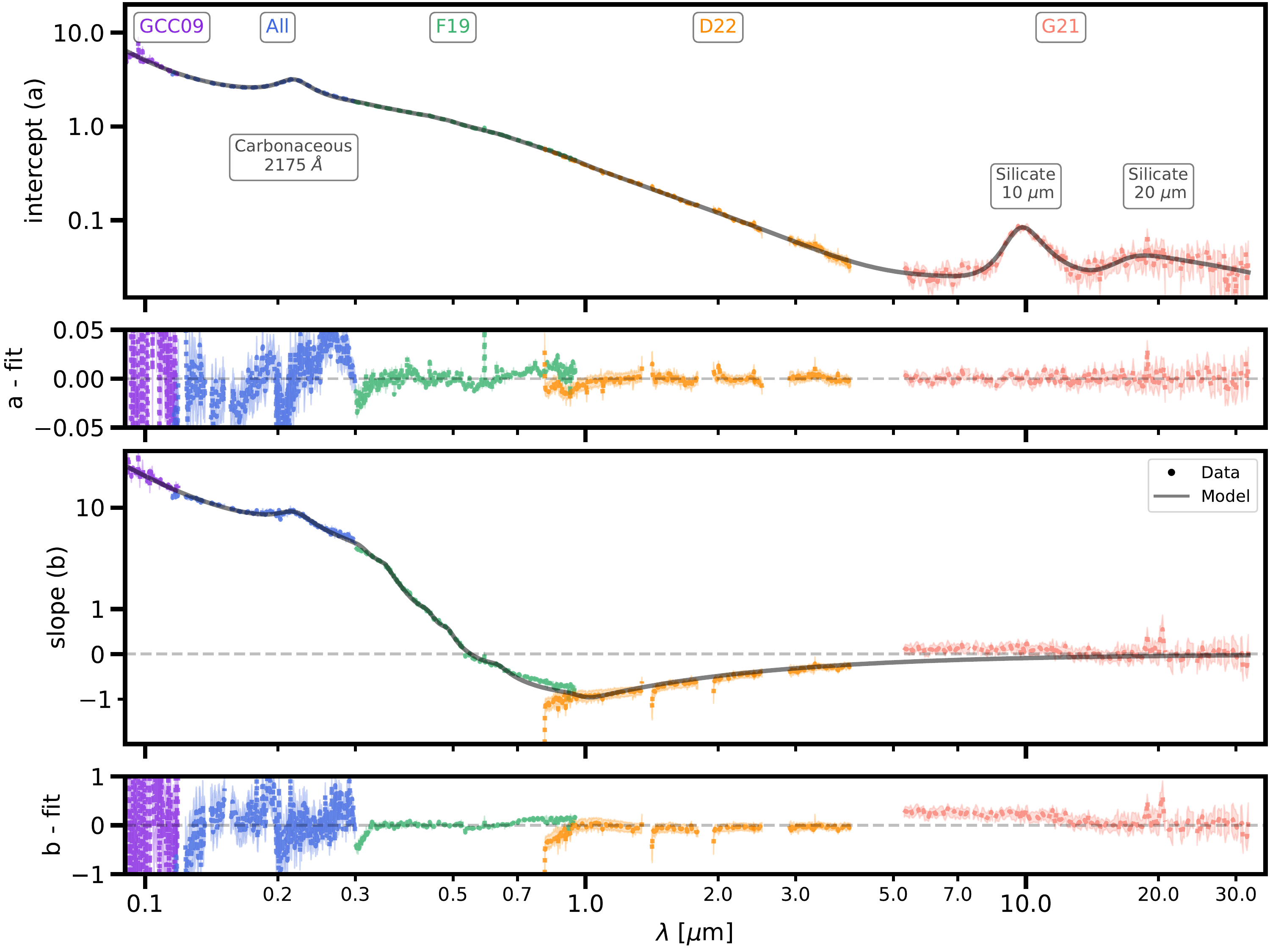}
\caption{Linear fit parameters for the full FUV to MIR wavelength range are plotted with the shaded regions giving the fit parameter uncertainties.
The results labeled as ``All" are from fits to the combined GCC09, F19, D22, and G21 extinction curves at UV wavelengths (1150--3100~\AA).
\label{fig:allwaves}}
\end{figure*}

The \rv\ relationship from the FUV to the MIR is shown in Fig.~\ref{fig:allwaves}.
The fits in the three wavelength regions (equations~\ref{eq:uv}, \ref{eq:opt}, \ref{eq:ir1}, and \ref{eq:ir2}) are combined in the overlap regions (0.3--0.33~\micron\ and 0.9--1.1~\micron) using the smoothstep function given in eq.~\ref{eq:smoothstep} with appropriate parameters.
Specifically, the coefficients of Eq.~\ref{eq_fitmod} are
\begin{equation}
\begin{matrix} a(\lambda) \\ b(\lambda) \end{matrix} =
\begin{cases}
\begin{matrix}
\left( \begin{matrix} a_\mathrm{uv} \\ b_\mathrm{uv} \end{matrix} \right)
\end{matrix}, & 0.0912 \mathrm{-} 0.3 \\
(1 - W1) \left( \begin{matrix} a_\mathrm{uv} \\ b_\mathrm{uv} \end{matrix} \right) + W1 \left( \begin{matrix} a_\mathrm{opt} \\ b_\mathrm{opt} \end{matrix} \right), & 0.3\mathrm{-}0.33 \\
\begin{matrix}
\left( \begin{matrix} a_\mathrm{opt} \\ b_\mathrm{opt} \end{matrix} \right)
\end{matrix}, & 0.33 \mathrm{-} 0.9 \\
(1 - W2) \left( \begin{matrix} a_\mathrm{opt} \\ b_\mathrm{opt} \end{matrix} \right) + W2 \left( \begin{matrix} a_\mathrm{ir} \\ b_\mathrm{ir} \end{matrix} \right), & 0.9\mathrm{-}1.1 \\
\begin{matrix}
\left( \begin{matrix} a_\mathrm{ir} \\ b_\mathrm{ir} \end{matrix} \right)
\end{matrix}, & 1.1 \mathrm{-} 32 \\
\end{cases}
\end{equation}
where the wavelength ranges are given in \micron\ and
\begin{eqnarray}
W1 & = & W(\lambda, 0.315~\micron, 0.03~\micron) \\
W2 & = & W(\lambda, 1.0~\micron, 0.2~\micron) .
\end{eqnarray}

The model and residuals between the data and the fits are shown in Fig.~\ref{fig:allwaves}.
As expected, the dust extinction at $R(V) = 3.1$ (intercept) shows generally decreasing extinction with increasing wavelength and strong dust features at \fbump, 10~\micron, and 20~\micron.
The \rv\ dependence (slope) is positive at wavelengths shorter than 0.55~\micron\ and negative for longer wavelengths where the change is due to normalizing the extinction curves by \av.
Comparing the intercept and slope plots, the \fbump\ feature clearly shows a dependence on \rv\ while the 10 and 20~\micron\ features do not.
This is expected as the \fbump\ feature is only seen for small carbonaceous grains while the 10 and 20~\micron\ features are due to silicate grains of all sizes \citep[e.g.,][]{Weingartner01, Zubko04, Jones13}.

\section{Discussion}
\label{sec:discussion}

\subsection{Comparison with previous R(V) relationships}

\begin{figure*}[tbp]
\epsscale{1.1}
\plotone{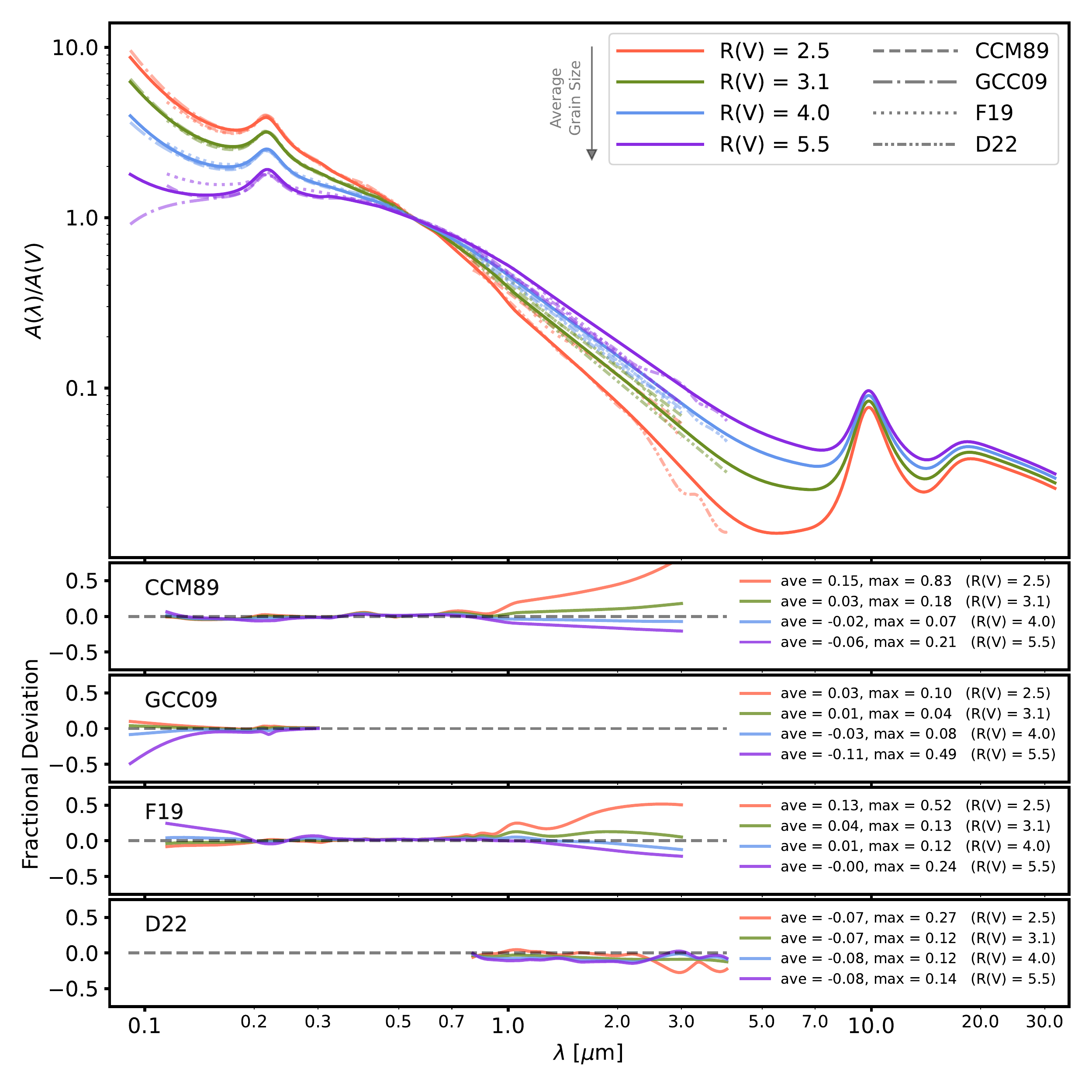}
\caption{Our \rv\ relationship evaluated for four representative \rv\ values is compared with literature \rv\ relationships for the same \rv\ values \citep{Cardelli89, Gordon09FUSE, Fitzpatrick19, Decleir22}.
The fractional deviations of the literature relationships from our \rv\ relationship are shown in the 4 bottom panels (e.g., (literature - ours)/ours).
The average of the deviations and maximum of the absolute value of the deviations at each \rv\ are given in each panel.}
\label{fig:selectrv}
\end{figure*}

Direct comparison of our and previous \rv\ relationships is not straightforward as previous \rv\ relationships have been based on fitting \alav\ versus $R(V)^{-1}$ \citep[e.g.,][]{Cardelli89, MaizAppellaniz14, Valencic04, Gordon09FUSE} and \elvebv\ versus \rv\ \citep[e.g.,][]{Fitzpatrick07}.
In addition, some are versus $R(V) - 3.1$ instead of just \rv\ \citep[e.g.,][]{Fitzpatrick19}.
As such, the $a$ and $b$ coefficients from different studies are not directly comparable.
Given this, we compare our and previous relationships by evaluating all at the same value of \rv\ providing for direct comparison of the dust extinction for different \rv\ values.
Specifically our relationship at four \rv\ values is compared to four representative literature relationships \citep{Cardelli89, Gordon09FUSE, Fitzpatrick19, Decleir22} in Fig.~\ref{fig:selectrv}.
Overall there is good agreement between our results and the literature in the optical range as expected given all are normalized in the V band.
There are significant deviations seen in the ultraviolet and in the near-IR.

In the UV, we see good agreement between the relationships expect for $R(V) = 5.5$.
At this \rv\ we agree well with CCM89, are lower than F19, and are higher than GCC09.
The difference with GCC09 below $\sim$0.15~\micron\ where the GCC09 relationship turns over and ours does not is notable.
Such a turnover is not seen in any of the high \rv\ extinction curves in their and our samples.
This is due to GCC09 not accounting for the uncertainties in \rv\ in their analysis and we were able to reproduce this behavior when we did the same.
This illustrates the importance of accurately including the full correlated uncertainties in deriving the \rv\ relationship using a method like 2DCORR.

In the NIR, the D22 and our relationships show a larger variation with \rv\ than the other literature relationships especially at high \rv\ values.
This is possibly due to CCM89, GCC09, and F19 only having a few photometric points at these wavelengths where the D22 and our relationships are based on spectroscopic extinction curves with many more points.
In comparison to D22 where spline interpolation was used, our relationship is smoother and shows slightly smaller variation with \rv\ due to the requiring continuity with the optical and MIR extinction and the use of a simple analytic function for the slope behavior (see Sec.~\ref{sec:rv_ir}).

\subsection{Deviations from the R(V) relationship}
\label{sec_deciations}

\begin{figure*}[tbp]
\epsscale{1.1}
\plotone{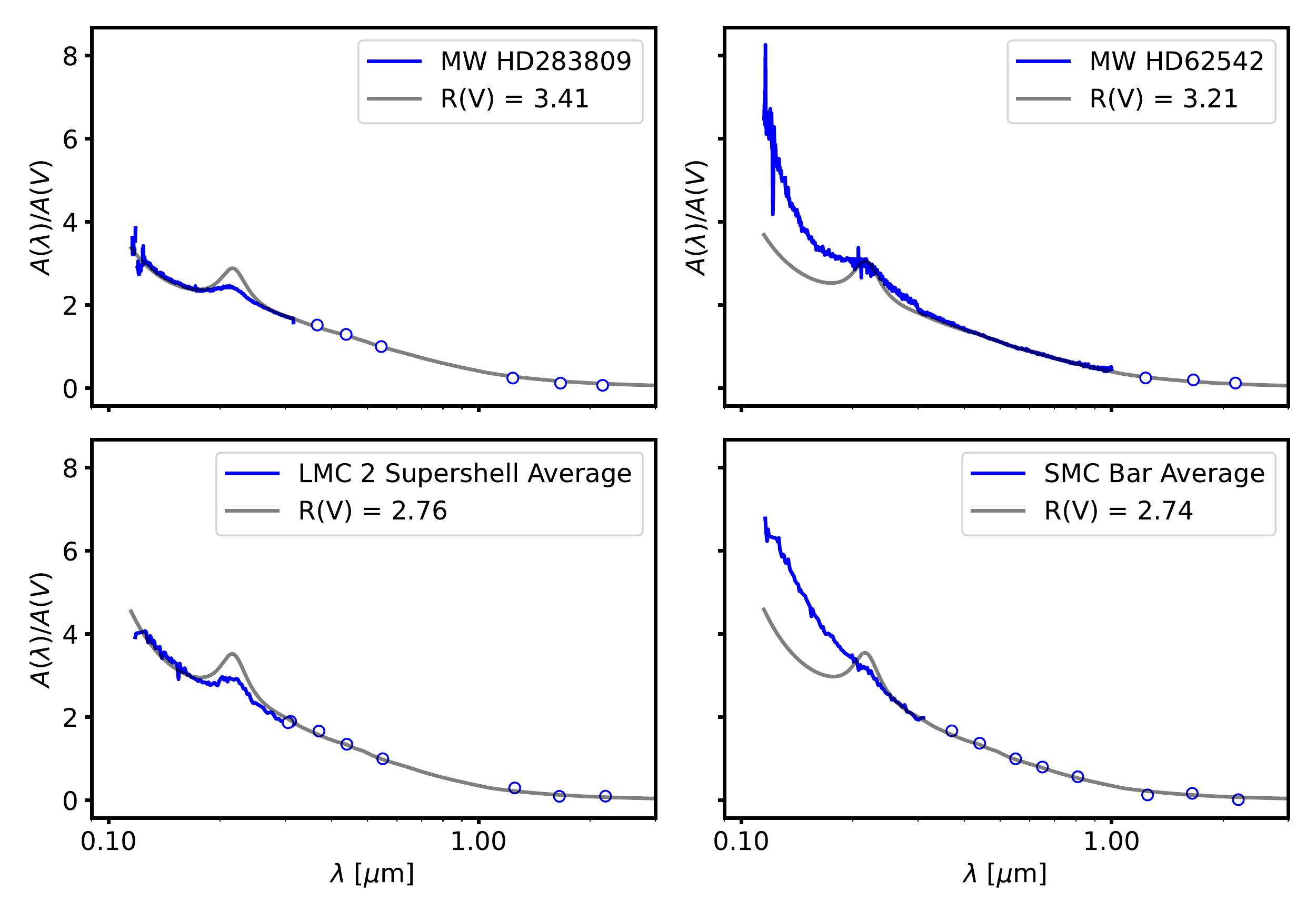}
\caption{Illustrating that not all extinction curves follow the \rv\ relationship, examples of curves that deviate are shown.
The MW sightlines are from this work and the LMC and SMC averages are from \citet{Gordon03}.
The main deviations are seen in the UV, both in weaker \fbump\ bumps and stronger far-UV rises.
\label{fig:deviations}}
\end{figure*}

The \rv\ relationship presented in this paper describes the average behavior as a function of \rv.
A small number of individual sightlines in the Milky Way have extinction curves that are significantly different than the average at their \rv\ as is illustrated in Fig.~\ref{fig:repwaves} at selected wavelengths.
Such deviations have been known for many years and are correlated between wavelengths \citep{Mathis92, Clayton00, Valencic03, Valencic04, Whittet04}.
In addition, there are sightlines in the Magellanic Clouds \citep{Gordon03} that show UV extinction curves that strongly deviate from \rv\ relationships.
This is illustrated in Fig.~\ref{fig:deviations} where selected sightlines in the Milky Way and averages in the Magellanic Clouds are shown along with their expected \rv\ dependent curves.
The HD~283809 sightline shows a \fbump\ bump that is significantly weaker than expected given its measured \rv, likely due to this sightline being dominated by dense material \citep{Whittet04}.
The HD~62542 sightline has a stronger far-UV and weaker \fbump\ bump than expected for its measured \rv, possibly due to processing from nearby hot stars \citep{Cardelli90}.
The weaker \fbump\ bump in the LMC~2 Supershell average is likely related to processing due to the 30~Dor star forming region \citep{Misselt99}.
Finally, the lack of a detectable \fbump\ bump and quite strong far-UV rise in the SMC Bar average is likely related to the low-metallicity, hard radiation field environment in this galaxy \citep{Gordon98}.
These deviations motivated \citet{Zagury07} and \citet{Gordon16} to propose different two parameter relationships where the Milky Way \rv\ relationship is mixed with a linear curve or the SMC Bar average.
In future work, we will investigate how well our updated \rv\ relationship works in this two parameter relationship to reproduce the average behavior of extinction curves throughout the Local Group.

\subsection{Linearity Assumption}
\label{sec:whylin}

Our and all other \rv\ dependent extinction relationships assume that \alav\ is linearly related to $R(V)^{-1}$.
Visual inspection of the top three plots for UV wavelengths in Fig.~\ref{fig:repwaves} shows that this assumption may be incorrect.
For example, there are significantly more points above the best fit line than below it for \rv\ values between 5 and 6.
We tested a quadratic model and found it was visually a bit better fit than a linear model for the UV wavelengths, but of marginal statistical significance.

Yet, this do raise the question: Is the assumption of linearity correct?
A simple thought experiment gives the answer: probably not.
The thought experiment is to imagine what would happen for \rv\ values significantly larger than 6.
For such \rv\ values, we would expect that the grain diameters would be large with respect to far-UV wavelengths and the grains to act more like bricks, just shadows with no wavelength dependence.
For such grains, the value of \alav\ for many wavelengths should be close to 1 as that is the expectation for bricks.
But if we extrapolate the linear relationships in the far-UV (see Fig.~\ref{fig:repwaves}), it is apparent that the values of \alav\ quickly drop below one even for \rv\ values just above 6.
Hence, the assumption of linearity is not theoretically supported and should break down at the shortest wavelengths first, just as is tentatively shown here.
This illustrates the need for increasing the sample of high \rv\ sightlines with measurements in the UV, especially in the far-UV.

\subsection{Use}

The \rv\ relationship determined in this work provides the average behavior of dust extinction as a function of \rv\ (i.e., average grain size) at spectroscopic resolution from 912~\AA\ to 32~\micron.
It can be reproduced from the analytic fits given in Sec.~\ref{sec:rv_uv}--\ref{sec:rv_ir} appropriately merged as discused in section \ref{sec:rv_all}.
In addition, this \rv\ relationship is included in the dust\_extinction python package package\footnote{https://github.com/karllark/dust\_extinction} \citep{dustextinction} as the G23 model.

\section{Summary \label{sec:summary}}

For the first time, we have derived a single \rv\ dependent extinction relationship from 912~\AA\ to 32~\micron\ at {\em spectroscopic} resolution using extinction measured along many Milky Way diffuse sightlines.
This provides a unified, empirical description of how dust extinction varies with \rv\ at high enough spectral resolution to resolve continuum variations and the major features.
All previous \rv\ relationships were based on a combination of spectroscopic and photometric measurements and/or did not cover the full wavelength range.
We used four existing representative samples of extinction curves to spectroscopically cover the far-UV \citep[0.0912--0.119~\micron,][]{Gordon09FUSE}, UV (0.115--0.31~\micron, all samples), optical \citep[0.3--0.95~\micron,][]{Fitzpatrick19}, NIR \citep[0.8--4.0~\micron,][]{Decleir22}, and MIR \citep[5--32~\micron,][]{Gordon21} wavelengths.
Linear fits of the normalized extinction at each wavelength versus \irv\ were done and the resulting linear fit coefficients as a function of wavelength were fit to analytic functions for each wavelength region.
Overall, we found good agreement in the \rv\ relationship in overlapping wavelength regions between the different samples with the small differences seen attributed to known calibration or extinction curve measurement issues.
This \rv\ relationship shows differences to previous relationships likely due to the full accounting for the correlated uncertainties in the extinction measurements and the constraint of a smooth relationship from the far-UV to the MIR.
Improvements to this \rv\ relationship are expected given planned observations with JWST in the NIR and MIR both expanding the sample and removing residual terrestrial atmospheric issues in the NIR.
In addition, new observations in the UV targeted at high \rv\ values are crucial to investigating the statistical significance of the signs of non-linearity seen.

The \rv\ relationship provides a summary of the average behavior of dust in the Milky Way as a function of the single variable \rv\ that is known to be a proxy for average grain size.
This \rv\ relationship can be used to study dust grain properties and to empirically account for the effects of dust on astrophysical sources extinguished by dust.
Given this relationship was determined versus \irv, the intercept gives the Milky Way average extinction curve at the average $R(V) = 3.1$.
The slope of the relationship shows how dust properties vary with \rv\ and clearly shows that the \fbump\ and three ISS optical features are dependent on grain size, but the 10 and 20~\micron\ silicate features are not.
We emphasize that the \rv\ dependent relationship gives the average extinction behavior versus wavelength and there are sightlines in the Milky Way and Magellanic Clouds that {\em} strongly deviate from the relationship.
Continued study of such deviant sightlines will likely provide important clues to the full nature of dust grains.
We expect this \rv\ relationship will be useful for many studies, especially those that combined observations at different wavelengths.
Such studies are becoming more common in this era where extensive multi-wavelength archives exist {\em and} new observations from the ground and space are possible.

The code used for the analysis and plots is available\footnote{https://github.com/karllark/fuv\_mir\_rv\_relationship} \citep{fuvmirrvrelationship}.
The data used in this paper is publically available \citep{fuvmirdata}.
The far-UV to MIR \rv\ dependent extinction relationship is available as the G23 model in the dust\_extinction python package\footnote{https://github.com/karllark/dust\_extinction}  \citep{dustextinction}.

\software{Astropy \citep{astropy:2013, astropy:2018, astropy:2022}; dust\_extinction \citep{dustextinction}; measure\_extinction \citep{measureextinction}}

\end{document}